\documentclass[aps,pra,superscriptaddress,twocolumn,a4paper,reprint,noeprint]{revtex4-1}
\usepackage[utf8]{inputenc}
\usepackage{graphicx}
\usepackage{amsmath}
\usepackage{amsfonts}
\usepackage{amssymb}
\usepackage{bbold}
\usepackage[usenames,dvipsnames,svgnames,table]{xcolor}
\usepackage[USenglish]{babel}
\usepackage{hyperref}
\usepackage{grffile}

\begin{document}

\title{Stability of the Weyl-semimetal phase on the pyrochlore lattice}

\author{Christoph Berke}
\email{berke@thp.uni-koeln.de}
\affiliation{Institute of Theoretical Physics, Technische Universit\"at Dresden, 01062 Dresden, Germany}
\affiliation{Institute for Theoretical Physics, University of Cologne, 50937 Cologne, Germany}

\author{Paolo Michetti}
\affiliation{Institute of Theoretical Physics, Technische Universit\"at Dresden, 01062 Dresden, Germany}

\author{Carsten Timm}
\email{carsten.timm@tu-dresden.de}
\affiliation{Institute of Theoretical Physics, Technische Universit\"at Dresden, 01062 Dresden, Germany}

\date{December 14, 2017}

\begin{abstract}

Motivated by the proposal of a Weyl-semimetal phase in pyrochlore iridates, we consider a Hubbard-type model on the pyrochlore lattice. To shed light on the question as to why such a state has not been observed experimentally, its robustness is analyzed. On the one hand, we study the possible phases when the system is doped. Magnetic frustration favors several phases with magnetic and charge order that do not occur at half filling, including additional Weyl-semimetal states close to quarter filling. On the other hand, we search for density waves that break translational symmetry and destroy the Weyl-semimetal phase close to half filling. The uniform Weyl semimetal is found to be stable, which we attribute to the low density of states close to the Fermi energy.

\end{abstract}

\maketitle

\section{Introduction}
\label{sec.intro}

Topological Weyl semimetals (TWSs) are currently receiving a lot of attention \cite{TuV13,VaV14,CTS16,YaF17,AMV17}.
At Weyl points, two otherwise nondegenerate bands touch with, to leading order, linear dispersion. In systems with both inversion and time-reversal symmetry (TRS), all bands are at least twofold degenerate so that Weyl points cannot occur. Hence, TWSs require inversion (parity) or time-reversal symmetry to be broken. Weyl states are possible if both are broken, in which case the Weyl points generically lie at different energies \cite{ZWB12}. Weyl states with broken inversion symmetry have been predicted \cite{HXB15,WFF15} and experimentally confirmed \cite{XBA15,LWY15,LWF15,XAB15,YLS15,LXW15,LYS16} for the family of compounds TaAs, NbAs, TaP, and NbP. Except for the so-called type-II TWSs \cite{SGW15}, these remain the only clearly established examples to date.

The second type of TWS, with broken time-reversal symmetry, has been proposed for pyrochlore iridates $R_2\mathrm{Ir}_2\mathrm{O}_7$ ($R$ is a rare-earth element) such as $\mathrm{Y}_2\mathrm{Ir}_2\mathrm{O}_7$ \cite{WTV11,WiK12,ChH12} and cobalt-based Heusler compounds $X\mathrm{Co}_2Z$ \cite{WVK16}. Recently, Weyl points have been predicted for the antiferromagnetic materials $\mathrm{Mn}_3\mathrm{Ge}$ and $\mathrm{Mn}_3\mathrm{Sn}$ \cite{YSZ17}.
Recent thermal and thermoelectric transport experiments on $\mathrm{Mn}_3\mathrm{Sn}$ \cite{LXD17} support the Weyl picture.
For these materials and for the proposed pyrochlore TWSs, magnetic frustration is essential for the Weyl state. 

Time-reversal symmetry can also be broken by a magnetic field. For example, $\mathrm{Bi}_{1-x}\mathrm{Sb}_x$ tuned to the transition between a topological and a trivial insulator at $x\approx 0.03$ has Dirac points, i.e., linear band touching points of twofold degenerate bands, which should split into pairs of Weyl points in a magnetic field \cite{KKW13}. There is also experimental evidence for Weyl points in the doped half-Heusler compound $\mathrm{GdPtBi}$ in a magnetic field \cite{HKW16}. The field leads to the splitting of the quadratic $\Gamma_8$ band-touching point into linearly dispersing Weyl points.

While there is still hope to realize a pyrochlore TWS by careful substitution, it is of interest why such a state is difficult to achieve. In this paper, we study the robustness of the Weyl state for pyrochlore iridates, addressing two issues: First, we analyze the possible phases of electrons on a pyrochlore lattice with local repulsive interactions when the system is doped. Note that we call a phase a TWS whenever it has band-touching points of linearly dispersing, nondegenerate bands, even if these points do not appear at the Fermi energy.
In comparison to a metal or a semimetal with quadratic band-touching point, the formation of Weyl points results in a shift of density of states away from the energy at which they appear, leading to a reduction of the free energy. The reduction is maximal if the Weyl points appear at the chemical potential. Hence, one expects the TWS phase to be disfavored by doping. We indeed find that the TWS phase becomes unstable towards other magnetic and nonmagnetic phases upon doping. These include phases with unequal electronic occupation numbers of different basis sites that nevertheless do not break translational symmetry. Since they do break other lattice symmetries, we still call them charge density waves (CDWs).

Second, we consider the stability of the undoped and weakly doped material with respect to density waves (DWs) that do break translational symmetry. A TWS can be unstable towards the formation of a DW due to the following mechanism \cite{WCA12,WaZ13,WaY16,CTS16,LPT16}:
A (commensurate) DW enlarges the unit cell and thus leads to backfolding of bands. This can place Weyl points of opposite chirality that are far apart in the original Brillouin zone close together. It is then possible to move them to the same $\mathbf{k}$ point by a small change of the mean-field order parameters, where they can annihilate and gap out. Thereby, density of states is shifted away from the chemical potential, pushing the energy of occupied states down and lowering the free energy. This reduction could overcompensate the increase incurred by changing the order parameters away from their optimum values for uniform states.
Laubach \textit{et al.}\ \cite{LPT16} have studied a model with tetragonal symmetry within the variational cluster approach and indeed find a fully gapped CDW for strong attractive interactions.
Yang \textit{et al.}\ \cite{YLR11} have proposed that an applied magnetic field can induce a CDW in a TWS with ordering vector $\mathbf{Q}$ connecting two Weyl points.
We show that while this mechanism can, in principle, work for the cubic pyrochlore lattice, the Weyl points in the pyrochlore system are very robust against this type of instability. This is ultimately due to the low density of states close to the Fermi energy.
Our results are obtained within a Hartree-Fock approximation \cite{WGK13} that is unrestricted beyond fixing a commensurate DW ordering vector $\mathbf{Q}$, which can be zero.

The rest of this paper is organized as follows: In Sec.\ \ref{sec.model}, we introduce the model Hamiltonian and review its mean-field decoupling. In Sec.\ \ref{sec.results}, we present results, first for the doping dependence and then for the search for DWs. The paper concludes with a summary and outlook in Sec.~\ref{sec.summary}.

\section{Model and theory}
\label{sec.model}

Iridium in pyrochlore iridates has a $5d^5$ electronic configuration. The iridium \textit{d} electrons dominate the low-energy physics. The crystal field splits the \textit{d} orbitals into an $e_g$ doublet above a $t_{2g}$ triplet \cite{WCK14}, which contains the five electrons. The $t_{2g}$ triplet can be described by an effective angular momentum $L_\mathrm{eff}=1$. Spin-orbit coupling splits the $t_{2g}$ levels into a doublet with effective total angular momentum $J_\mathrm{eff}=1/2$ above a quartet with $J_\mathrm{eff}=3/2$. 
In the description of the low-energy physics, we can ignore both the completely filled quartet and the empty $e_g$ orbitals, which are relatively well separated from the Fermi energy \cite{WGK13,WCK14}. We are left with an effective Kramers doublet since the two basis states are related by time reversal. They will be distinguished by an effective-spin index $\sigma = {\uparrow},{\downarrow}$. The noninteracting part of the Hamiltonian is written as
\begin{equation}
H_0 = \sum_{ij} \sum_{\sigma\sigma'} c_{i\sigma}^\dagger h^{\sigma\sigma'}_{ij} c_{j\sigma'} ,
\label{model.H0.2}
\end{equation}
where $i$ and $j$ enumerate all iridium sites, which form a pyrochlore structure. The underlying Bravais lattice is face-centered cubic (fcc). We take the size of the conventional nonprimitive fcc unit cell to be $a$. Then the positions of iridium ions can be taken to be $(0,0,0)$, $a/4\,(0,1,1)$, $a/4\,(1,0,1)$, and $a/4\,(1,1,0)$, which form a tetrahedron. The explicit Hamiltonian turns out to be simpler if we shift the origin to the center of the tetrahedron. Setting $a=4$, we obtain the basis vectors
\begin{align}
\mathbf{b}_1 &= (-1/2,-1/2,-1/2) , \\
\mathbf{b}_2 &= (-1/2,+1/2,+1/2) , \\
\mathbf{b}_3 &= (+1/2,-1/2,+1/2) , \\
\mathbf{b}_4 &= (+1/2,+1/2,-1/2) .
\end{align}
In the absence of magnetic or charge order, the space group is $\mathrm{Fd\bar{3}m}$ (227) with point group $O_h$.

The coefficients $h^{\sigma\sigma'}_{ij}$ in Eq.\ (\ref{model.H0.2}) are constrained by lattice symmetries and time-reversal symmetry. Including nearest and next-nearest neighbor hopping, the most general symmetry-allowed Hamiltonian reads \cite{WGK13}
\begin{align}
H_0 &= \sum_{\langle ij\rangle} c_i^\dagger (t_1 + it_2\,
  \mathbf{d}_{ij}\cdot\mbox{\boldmath$\sigma$}) c_j \nonumber \\
& {}+ \sum_{\langle\langle ij\rangle\rangle} c_i^\dagger \big[t_1'
  + i(t_2'\, \mathbf{R}_{ij} + t_3'\, \mathbf{D}_{ij}) \cdot\mbox{\boldmath$\sigma$}\big]
  c_j .
\label{model.H0.4}
\end{align}
Here, $c_i$ is a spinor formed by the annihilation operators $c_{i\uparrow}$ and $c_{i\downarrow}$, $\mbox{\boldmath$\sigma$}$ is the vector of Pauli matrices, and
\begin{align}
\mathbf{d}_{ij} &= 2 \mathbf{b}_i \times \mathbf{b}_j , \\
\mathbf{R}_{ij} &= (\mathbf{b}_i-\mathbf{b}_k) \times (\mathbf{b}_k-\mathbf{b}_j) , \\
\mathbf{D}_{ij} &= \mathbf{d}_{ik} \times \mathbf{d}_{kj} ,
\end{align}
where $\mathbf{b}_i$ is understood as the basis vector belonging to site $i$ in the pyrochlore structure and in the last two expressions site $k$ is the common nearest neighbor of sites $i$ and $j$. This Hamiltonian is identical to the one studied by Witczak-Krempa \textit{et al.}\ \cite{WGK13}, but with the characteristic vectors written in a slightly simpler form.

Using the Slater-Koster method \cite{SlK54}, the hopping parameters in Eq.\ (\ref{model.H0.4}) can be expressed in terms of microscopic hopping amplitudes between iridium ions directly through $\sigma$ and $\pi$ bonds ($t_\sigma$, $t_\sigma'$, $t_\pi$, $t_\pi'$, where the prime denotes next-nearest-neighbor hopping) and via oxygen ions ($t_\mathrm{O}$) \cite{PeB10,WiK12,WGK13}:
\begin{align}
t_1 &= \frac{130}{243}\, t_\mathrm{O} + \frac{17}{324}\, t_\sigma - \frac{79}{243}\, t_\pi , \\
t_2 &= \frac{28}{243}\, t_\mathrm{O} + \frac{15}{243}\, t_\sigma - \frac{40}{243}\, t_\pi , \\
t_1' &= \frac{233}{2916}\, t_\sigma' - \frac{407}{2187}\, t_\pi' , \\
t_2' &= \frac{1}{1458}\, t_\sigma' + \frac{220}{2187}\, t_\pi' , \\
t_3' &= \frac{25}{1458}\, t_\sigma' + \frac{460}{2187}\, t_\pi' .
\end{align}
The hopping parameters are chosen as in Ref.\ \cite{WGK13}, to facilitate comparison:
\begin{align}
t_\mathrm{O} &= 1 , \\
t_\pi &= -\frac{2}{3}\, t_\sigma , \\
t'_\sigma &= 0.08\, t_\sigma , \\
t'_\pi &= 0.08\, t_\pi .
\end{align}
The iridium-iridium hopping is parameterized in terms of $t_\sigma$, the direct nearest-neighbor hopping amplitude along $\sigma$ bonds in units of the hopping via oxygen.

The full Hamiltonian is obtained by adding a local Hubbard repulsion,
\begin{equation}
H = H_0 + U \sum_i c_{i\uparrow}^\dagger c_{i\uparrow} c_{i\downarrow}^\dagger c_{i\downarrow} ,
\end{equation}
where the repulsion energy $U$ is also given in units of $t_\mathrm{O}$. Hydrostatic pressure \cite{TIM12} is expected to affect the overall hopping scale $t_\mathrm{O}$ more strongly than the ratios $t_\sigma\equiv t_\sigma/t_\mathrm{O}$ etc. Thus its leading effect in our model is to tune $U\equiv U/t_\mathrm{O}$.
For large $U$, the ground state is a Mott insulator. The Hubbard interaction together with the spin-orbit coupling in $H_0$ lead to anisotropic magnetic interactions in the Mott phase \cite{JaK09}. However, for intermediate values of $U$, as appropriate for iridates, the system can be a metal or a semimetal. This required an electronic instead of a spin-only description.

The interaction term is decoupled in the Hartree-Fock approximation. The mean-field parameters $\langle c_{i\sigma}^\dagger c_{i\sigma'}\rangle$ are expressed in terms of the average occupations and spin polarizations at sites $i$,
\begin{align}
n_i &= \sum_\sigma \langle c_{i\sigma}^\dagger c_{i\sigma} \rangle , \\
\mathbf{m}_i &= \frac{1}{2} \sum_{\sigma\sigma'} \langle c_{i\sigma}^\dagger
  \mbox{\boldmath$\sigma$}_{\sigma\sigma'} c_{i\sigma'} \rangle .
\end{align}
We will first consider solutions that do not reduce the translational symmetry. Then, $n_i$ and $\mathbf{m}_i$ can only vary between the four basis sites. Since the total occupation $\sum_i n_i$ is fixed by the doping level, we are left with 15 mean-field parameters, which form the order parameters of the magnetic and charge order. To study DWs that break translational symmetry, we will consider supercells consisting of $n$ primitive unit cells. This leads to $16n-1$ order parameters. These parameters are calculated selfconsistently by iteration, using the bilinear Hartree-Fock Hamiltonian. Although we are interested in the ground states, we have used a small nonzero temperature (inverse temperatures $\beta=1/k_BT\ge 600$) to improve convergence. The chemical potential has been updated during the iteration to fix the total occupation. The iteration is started with random values for $n_i$ and $\mathbf{m}_i$ and usually on the order of ten independent runs are performed. If these converge to different selfconsistent solutions, indicating the existence of multiple metastable states, the solution with the lowest Hartree-Fock free energy is selected.

\section{Results}
\label{sec.results}

The main objectives in this section are to study the stability of the Weyl state upon doping and to search for DWs that break translational symmetry. To set the stage and to make contact with previous results, we first reobtain the phase diagram of Witczak-Krempa \textit{et al.}\ \cite{WGK13} for the undoped (half-filled) system. We will also present new results for this case. After that, we turn to the analysis of the doping dependence and finally to the search for DWs.

\begin{figure}[htb]
\begin{center}
\includegraphics[width=\columnwidth]{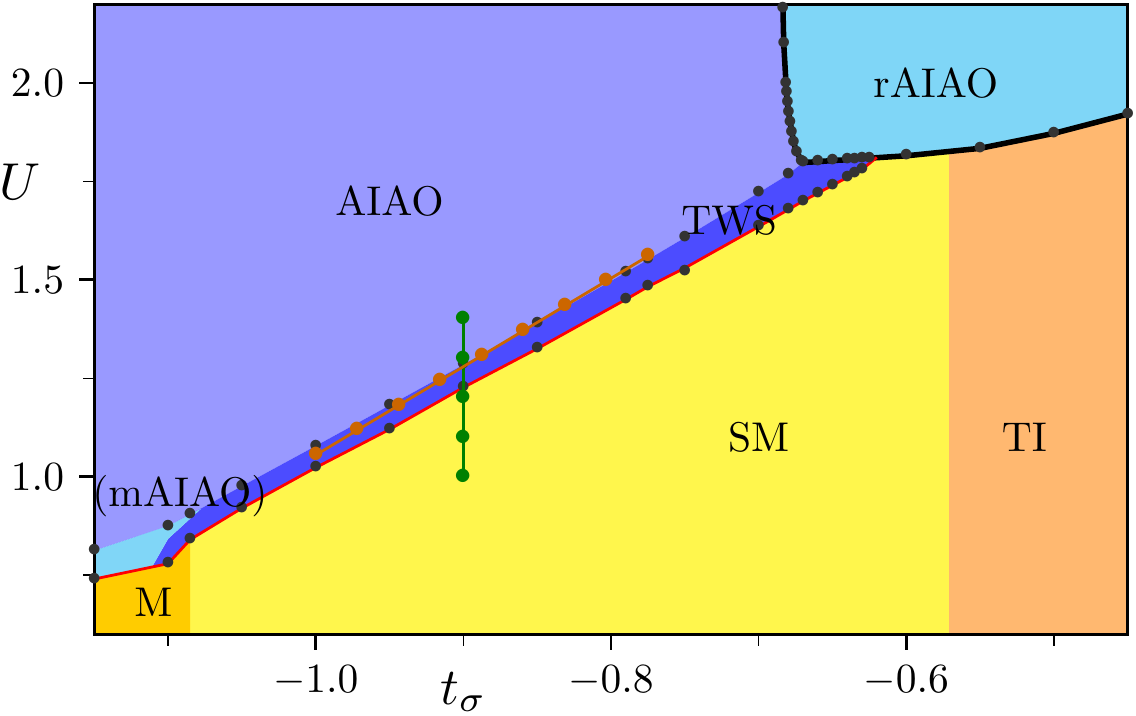}
\end{center}
\caption{Mean-field phase diagram for the undoped model. The precise location of the phase boundary has been determined at the black dots. In the blue regions, TRS is broken, in the yellow and orange ones it is preserved. The acronyms are explained in the text. The heavy black line corresponds to a first-order phase transition, all other transitions are continuous. The second-order transitions where magnetic order sets in are marked by thin red lines. The dark green and orange lines with dots indicate cuts through the phase diagram along which the effect of doping will be studied below.}
\label{fig.PD1}
\end{figure}

\subsection{Half-filled case}
\label{sec.undoped}

The mean-field phase diagram in the $t_\sigma$-$U$ plane for the undoped case is shown in Fig.\ \ref{fig.PD1}. The phase diagram agrees quantitatively with the one in Ref.\ \cite{WGK13} and also qualitatively with Ref.\ \cite{LBK13}. Recently, the pyrochlore material $\mathrm{Eu_2Ir_2O_7}$ has been studied by a very similar unrestricted Hartree-Fock approach for a tight-binding model by Wang \textit{et al.}\ \cite{WWF17}. In this case, the hopping amplitudes were obtained from density-functional calculations without spin-orbit coupling and the spin-orbit coupling and Hubbard-$U$ terms added afterwards. The phase diagram in the plane spanned by spin-orbit coupling strength and $U$ is obtained. The sequence of phases for increasing $U$ for the undistorted pyrochlore structure is similar to Fig.\ \ref{fig.PD1}. We return to the comparison below.

Two main regions can be distinguished: the yellow and orange areas for small $U$ denote phases without magnetic order. In the blue areas for large $U$, the magnetic moments are nonzero and TRS is broken. Mostly following Ref.\ \cite{WGK13}, the phases are denoted as metal (M), semimetal with quadratic band-touching point (SM), topological insulator (TI), metal with all-in-all-out order (mAIAO), topological Weyl semimetal (TWS), insulator with all-in-all-out magnetic order (AIAO), and insulator with rotated all-in-all-order order (rAIAO). All transitions are continuous except for the first-order transitions towards the rAIAO phase. The continuous transitions can be understood as Lifshitz transitions in the sense that they involve changes in the topology of the intersection of a band with the Fermi energy or of intersections between bands \cite{Vol07,Vol17}. It is, however, useful to single out second-order transitions at which the mean-field spin polarizations $\mathbf{m}_i$ start to deviate from zero. These transitions are highlighted by thin red lines in Fig.\ \ref{fig.PD1}.
The magnetic order and typical band structures for the various phases are discussed in Ref.\ \cite{WGK13}. We here focus on deviations from and additions to the previous results.

\begin{figure}[htb]
\begin{center}
\includegraphics[width=0.6\columnwidth]{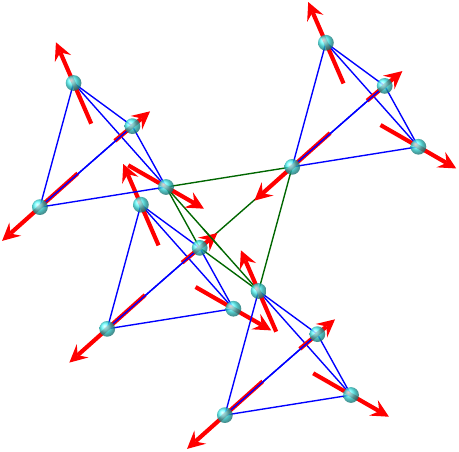}
\end{center}
\caption{Sketch of the all-in-all-out magnetic order in the TWS, mAIAO, and AIAO phases.}
\label{fig.sketch.AIAO}
\end{figure}

\begin{figure}[htb]
\begin{center}
\includegraphics[width=0.5\columnwidth]{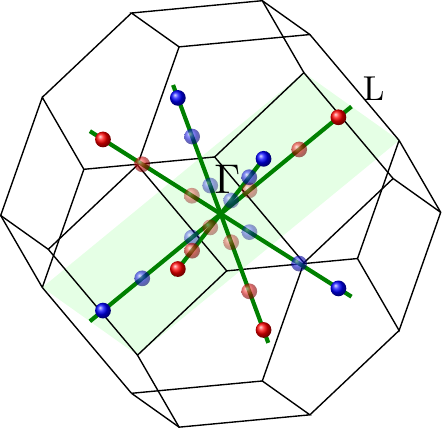}
\end{center}
\caption{Positions of the eight Weyl points in the fcc Brillouin zone. The Weyl points are shown for three different values of the spin polarization $|\mathbf{m}_i|$ distinguished by different degrees of transparency. The red and blue color corresponds to chirality $+1$ and $-1$, respectively.}
\label{fig.WP1}
\end{figure}

The TWS, mAIAO, and AIAO phases have the same magnetic all-in-all-out order. For a given tetrahedron, the spins at all four corners point either towards or away from the center, as sketched in Fig.\ \ref{fig.sketch.AIAO}. The order breaks TRS and also a $\mathbb{Z}_2$ (Ising) symmetry corresponding to inverting all spins. The occupation and the magnitude of the spin polarization are identical for all sites. The averages of the four spins are
\begin{align}
\mathbf{m}_1 &= \frac{\alpha}{2\sqrt{3}}\, (1,1,1) , \\
\mathbf{m}_2 &= \frac{\alpha}{2\sqrt{3}}\, (1,-1,-1) , \\
\mathbf{m}_3 &= \frac{\alpha}{2\sqrt{3}}\, (-1,1,-1) , \\
\mathbf{m}_4 &= \frac{\alpha}{2\sqrt{3}}\, (-1,-1,1) ,
\end{align}
with $0 < |\alpha|\le 1$.
The magnetic space group is $\mathrm{Fd\bar{3}m'}$ \cite{Lit13} with the magnetic point group $\mathrm{m\bar{3}m'} = O_h(T_h)$ \cite{DDJ08}.
The all-in-all-out order is the natural equilibrium state for spins on the frustrated pyrochlore lattice with strong local easy-axis anisotropy and antiferromagnetic interactions \cite{GWJ12,SMM13,SHT15}.

The electronic structure in the three phases is different. For large $U$, leading to large spin polarizations, the AIAO phase is an insulator. For smaller spin polarizations, there is a transition towards a TWS with eight Weyl points lying on the $\Lambda$ lines, which connect the $\Gamma$ and L points \cite{WGK13} (in the earlier Ref.\ \cite{WTV11}, 24 Weyl points were found). The Weyl points form the corners of a cube, as shown in Fig.\ \ref{fig.WP1}. At the transition from the TWS to the AIAO phase for increasing $U$, the Weyl points annihilate pairwise at the L points. At the transition to the SM phase for decreasing $U$, all eight Weyl points instead merge at $\Gamma$ and the spin polarizations $\mathbf{m}_i$ vanish continuously. The bands are not split at $\Gamma$ since the relevant bands transform according to the $\Gamma_8$ irrep of $O_h$. The ensuing nonmagnetic SM phase is thus a semimetal with a quadratic band-touching point \cite{MXK13}. The third, mAIAO region is analyzed further below.

For large $U$ and small $|t_\sigma|$, a rotated all-in-all-out (rAIAO) order is found. It can be constructed from the perfect all-in-all-out order by dividing the four spins of one tetrahedron into two pairs. The two spins in the same pair are rotated in the same plane, which contains the locations of the two spins and the center of the opposite tetrahedron edge, in opposite directions. The absolute value of the rotation angle is the same for all spins. If $\mathbf{m}_1$ and $\mathbf{m}_2$ are rotated in the same plane, the resulting average spins read
\begin{align}
\mathbf{m}_1 &= (a,b,b) ,
\label{half.rAIAO.m1} \\
\mathbf{m}_2 &= (a,-b,-b) , \\
\mathbf{m}_3 &= (-a,b,-b) , \\
\mathbf{m}_4 &= (-a,-b,b) .
\label{half.rAIAO.m4}
\end{align}
The net magnetization still vanishes. The all-in-all-out order is the special case $a=b=\alpha/2\sqrt{3}$. The rotated all-in-all-out order also lowers the symmetry to tetragonal. Since there are three ways to split four spins into two pairs and the free energy is invariant under inverting all spins, there are six degenerate equilibrium states.

The rAIAO phase is the same one that Wang \textit{et al.}\ \cite{WWF17} find at large $U$ for the ideal pyrochlore structure and denote by AF2. This can be seen by comparing Eqs.\ (\ref{half.rAIAO.m1})--(\ref{half.rAIAO.m4}) to the spins given in Fig.\ 1(b) in Ref.\ \cite{WWF17}. We conclude that the tight-binding model of Ref.\ \cite{WWF17} corresponds to small $|t_\sigma|$ in our phase diagram.

\begin{figure}[htb]
\begin{center}
\includegraphics[width=0.9\columnwidth]{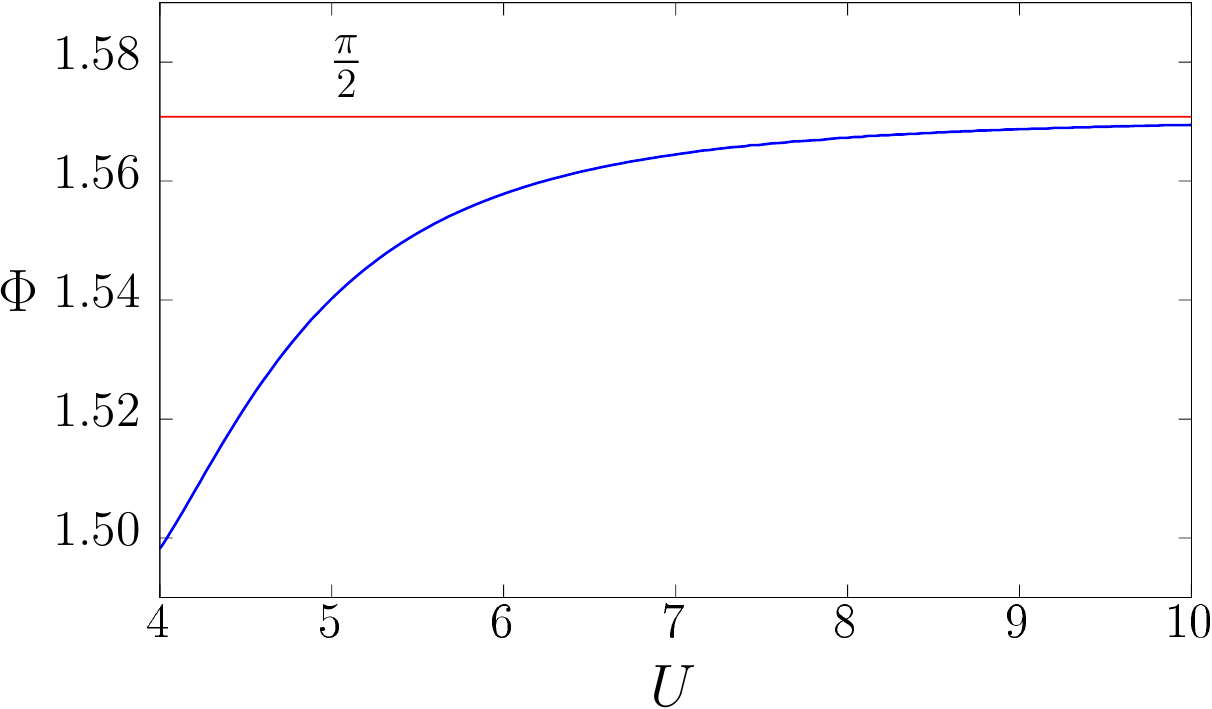}
\end{center}
\caption{Rotation angle $\Phi$ of the average spins in the rAIAO phase relative to the AIAO phase, as a function of the interaction strength $U$. The hopping scale is set to $t_\sigma=-0.5$.}
\label{fig.rAIAO.angle}
\end{figure}

Our results differ from Refs.\ \cite{WiK12,WGK13,WWF17} in that the rotation angle always equals $\pi/2$ in these works. Instead, we find the angle to vary, as shown in Fig.\ \ref{fig.rAIAO.angle}. In the limit of large $U$, the rotation angle approaches $\pi/2$. This limit is consistent with Monte Carlo simulations for a pure spin model with Dzyaloshinski-Moriya interaction~\cite{ECS05}.

\begin{figure}[htb]
\begin{center}
\includegraphics[width=\columnwidth]{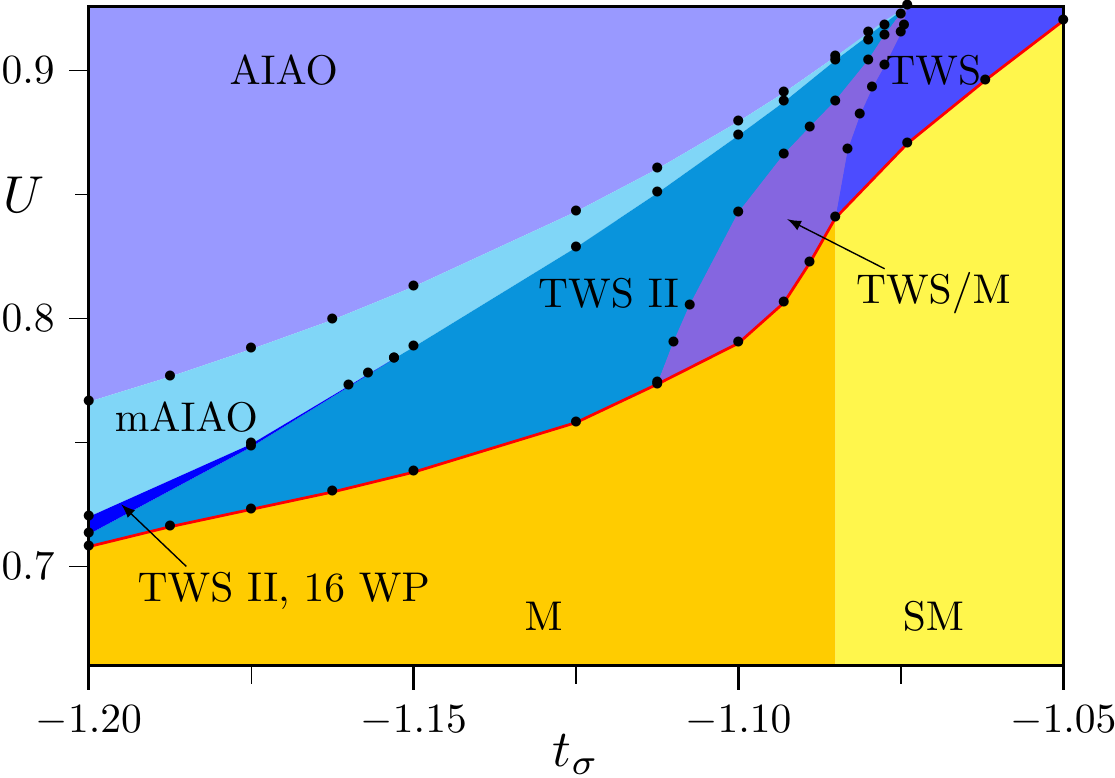}
\end{center}
\caption{Close up of the metallic all-in-all-out (mAIAO) region of the phase diagram Fig.\ \ref{fig.PD1}. The precise location of the phase boundary has been determined at the black dots. The acronyms are explained in the text.}
\label{fig.PD2}
\end{figure}

The region denoted as metallic all-in-all-out (mAIAO) in Fig.\ \ref{fig.PD1} and as metallic antiferromagnet in Ref.\ \cite{WGK13} is particularly interesting. Figure \ref{fig.PD2} reveals that several distinct phases are present. These are best understood by starting from the TWS phase and increasing $|t_\sigma|$. This pushes the conduction band down away from the $\Gamma$ point and leads to the appearance of electron-type Fermi pockets at the L points. The chemical potential is then shifted downward from the Weyl points into the valence band. The result is a compensated metal with Weyl points, denoted by TWS/M. For increasing $|t_\sigma|$, the Weyl cones are also tilted so that the dispersion in one direction eventually becomes flat and then changes sign. The result is a type-II TWS as introduced by Soluyanov \textit{et al.}\ \cite{SGW15}. This phase is denoted by TWS~II.

Increasing $|t_\sigma|$ further, the Weyl points move to the L points and annihilate pairwise. The resulting phase does not have any band-touching points but is a compensated metal due to the overlap in energy of bands in different parts of the Brillouin zone. It retains all-in-all-out order and is denoted by mAIAO in Fig.\ \ref{fig.PD2}. However, at large $|t_\sigma|$ and small $U$, something else happens: at the four L points, four additional pairs of Weyl points are formed. The system then has 16 Weyl points, all of which are of type II. This phase is denoted by TWS~II, 16 WP. The newly created and the old Weyl points then move towards each other for increasing $|t_\sigma|$ or $U$ and finally annihilate pairwise, again leading to the mAIAO phase.

For smaller $U$, we instead start in the SM phase with quadratic band touching at $\Gamma$. For increasing $|t_\sigma|$, the band structure deforms similarly to what happens in the TWS phase. Fermi pockets appear at the L points and the chemical potential is shifted into the valence band. The resulting compensated metal is denoted by M.

\subsection{Doping dependence}
\label{sec.doping}

In the following, we consider phases along two representative cuts through the phase diagram, which are indicated in Fig.\ \ref{fig.PD1} by solid lines with dots. As the second control parameter we take the average number $\rho$ of electrons in the $J_\mathrm{eff}=1/2$ doublet per iridium so that $\rho=1$ corresponds to half filling, i.e., the undoped case considered previously. As noted above, doping away from half filling should shift the chemical potential away from the Weyl points in the TWS phase and thus reduce the free energy gain that stabilizes the TWS state. In this section, we only consider states that do not break translational symmetry.

\begin{figure}[htb]
\begin{center}
\includegraphics[width=\columnwidth]{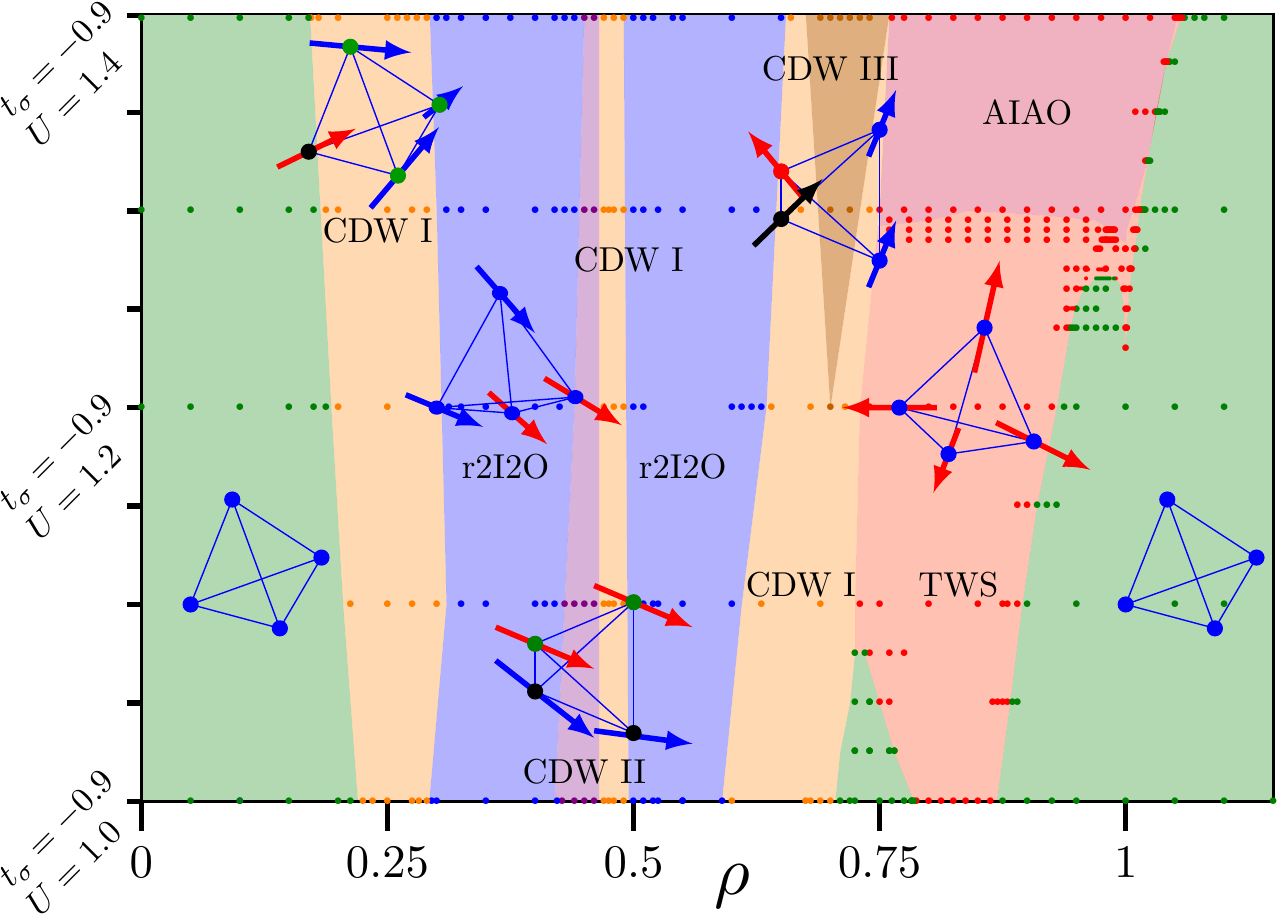}
\end{center}
\caption{Mean-field phase diagram for $t_\sigma=-0.9$ and varying interaction strength $U$ and filling $\rho$. The considered parameter points are shown as colored dots. The regions shaded in less saturated colors represent our best guess at the regions occupied by the various phases. At $\rho=1$, this phase diagram corresponds to the dark green line with dots in Fig.\ \ref{fig.PD1}. The magnetic and charge order in the various phases is indicated in the insets. The acronyms are explained in the text.}
\label{fig.PDcut1}
\end{figure}

We first consider a cut along the dark green line with dots in Fig.\ \ref{fig.PD1}, for which $t_\sigma=-0.9$ is fixed and $U$ is varied between $1.0$ and $1.4$. The phase diagram is shown in Fig.\ \ref{fig.PDcut1}. Several distinct magnetic and also charge orders are found, which are illustrated by the insets. For electron doping, $\rho>1$, magnetic order vanishes rapidly. In case of the gapped AIAO phase, this is easy to understand. The AIAO order splits the band touching at the $\Gamma$ point, regardless of the Fermi energy. For our parameters, the effective mass of the conduction band at the $\Gamma$ point is \emph{larger} than that of the valence band. Consequently, the density of states close to the energy of the touching point is asymmetric, being larger in the conduction band. The opening of a gap is then unfavorable for electron doping because the energy increase of occupied conduction-band states overwhelms the energy reduction of valence-band states. This can only be overcome by a stronger interaction $U$, which is why the transition shifts to larger $U$ for increasing electron doping. The argument is given in more detail in Appendix \ref{app.phasym}. The situation is not fundamentally different for the TWS phase, except that no full gap is opened but density of states is still shifted away from the Weyl points. We have not found any phases other than the nonmagnetic semimetal for larger $\rho$.

\begin{figure}[htb]
\begin{center}
\includegraphics[width=0.45\columnwidth]{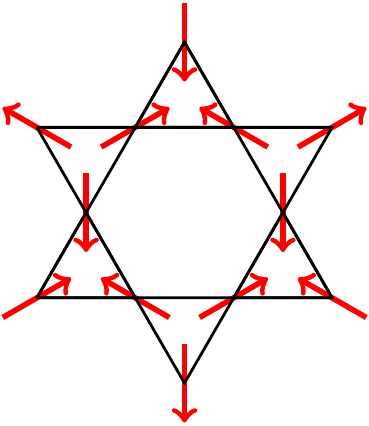}
\end{center}
\caption{Projections of the spin in a kagome layer in the CDW~I phase onto the plane of that layer.}
\label{fig.CDW.I.kagome}
\end{figure}

In the following, we discuss the phases for fillings $0<\rho\lesssim 1$, starting from small $\rho$. It should be noted that for a realistic description of a pyrochlore material in this doping range, additional orbitals, in particular the $J_\mathrm{eff}=3/2$ multiplet need to be taken into account.
For small $\rho$, the ground state is nonmagnetic. Upon increasing $\rho$, we find a continuous phase transition towards the state marked CDW~I in Fig.\ \ref{fig.PDcut1}. This phase has ferrimagnetic order with magnetization along the (111) or an equivalent direction. The average spins take the form
\begin{align}
\mathbf{m}_1 &= (a,a,a) ,
\label{dope.CDW.I.m1}
\\
\mathbf{m}_2 &= (b,c,c) , \\
\mathbf{m}_3 &= (c,b,c) , \\
\mathbf{m}_4 &= (c,c,b) ,
\label{dope.CDW.I.m4}
\end{align}
note the arrows in the corresponding inset. Since the spins at sites 2, 3, and 4 are related to each other by lattice symmetries, but not to the spin at site 1, the occupation number of site 1 is generically different from the others, $n_1\neq n_2=n_3=n_4$ (indicated by different colors of the dots in the inset). Hence, this is a CDW state, which, however, does not break translational symmetry.
The order is most easily understood by viewing the pyrochlore lattice as an alternating stack of triangular and kagome lattices. In the (111) direction, the triangular layers consist solely of sites with number 1 and the kagome layers of sites 2, 3, and 4. The spins in the triangular layers point towards one tetrahedron center and away from the other. The spins in the kagome layers are roughly parallel to the ones in the triangular layers but tilted towards and away from the centers of adjacent triangles of the kagome lattice. The projections of these spins onto the plane of the kagome layer are shown in Fig.\ \ref{fig.CDW.I.kagome}. The state can also be understood as a ferromagnet that is perturbed by local Ising anisotropies. Since there are four ways to decompose the lattice into triangular and kagome layers and the energy is invariant under inversion of all spins, the equilibrium state is eightfold degenerate.

Increasing $\rho$ further, we find a transition to a state with average spins
\begin{align}
\mathbf{m}_1 &= (a,b,b) ,
\label{dope.r2I2O.m1} \\
\mathbf{m}_2 &= (a,-b,-b) , \\
\mathbf{m}_3 &= (a,-b,b) , \\
\mathbf{m}_4 &= (a,b,-b)
\label{dope.r2I2O.m4}
\end{align}
(or other states related by symmetries) and equal occupation numbers. The state is ferrimagnetic with magnetization along (100) or equivalent directions. Note that the spin averages differ from the rAIAO order in Eqs.\ (\ref{half.rAIAO.m1})--(\ref{half.rAIAO.m4}) only in the signs of $\mathbf{m}_3$ and $\mathbf{m}_4$. Hence, the new configuration is obtained by the same pairwise rotations as the rAIAO order but starting from a state with two spins pointing in and two pointing out of the tetrahedron. We call the new configuration rotated two-in-two-out (r2I2O) order. In the studied parameter range, we always find $|a|\gg |b|$, i.e., the state is close to a collinear ferromagnet.
Since there are six possibilities to choose the in-pointing two spins, the degeneracy is sixfold.

Next, there is another transition to a state with
\begin{align}
\mathbf{m}_1 &= \mathbf{m}_2 = (0,a,-a) , \\
\mathbf{m}_3 &= (c,b,-b) , \\
\mathbf{m}_4 &= (-c,b,-b) .
\end{align}
As suggested by symmetry, the occupation numbers satisfy $n_1=n_2\neq n_3=n_4$. The first two occupations are smaller than the latter two. Hence, this is another CDW state that does not break translational symmetry. Such a charge order, without the magnetic order, has also been found by Kurita \textit{et al.}\ \cite{KYI11}. We denote this state by CDW~II. The first two spins are oriented parallel to the tetrahedron edge connecting sites 3 and 4. The other two spins are nearly parallel to the first two ($a$ and $b$ have the same sign) but canted in opposite directions. The state is ferrimagnetic with magnetization along ($01\bar{1}$) or equivalent directions, i.e., along one of the tetrahedron edges. Since there are six possibilities to choose the two equal spins and the energy is invariant under inversion of all spins, the equilibrium state is twelvefold degenerate.

Close to quarter filling ($\rho=0.5$), the order CDW~I recurs but the values of $a$, $b$, and $c$ in Eqs.\ (\ref{dope.CDW.I.m1})--(\ref{dope.CDW.I.m4}) are very similar, i.e., this state is close to a collinear ferromagnet. At quarter filling, Kurita \textit{et al.}\ \cite{KYI11} have found a CDW with three charge-poor sites, i.e., $n_1>n_2=n_3=n_4$ and without magnetic order for large nearest-neighbor Coulomb repulsion. For dominating on-site repulsion $U$, they find a ferromagnet.

\begin{figure}[htb]
\begin{center}
\includegraphics[width=0.9\columnwidth]{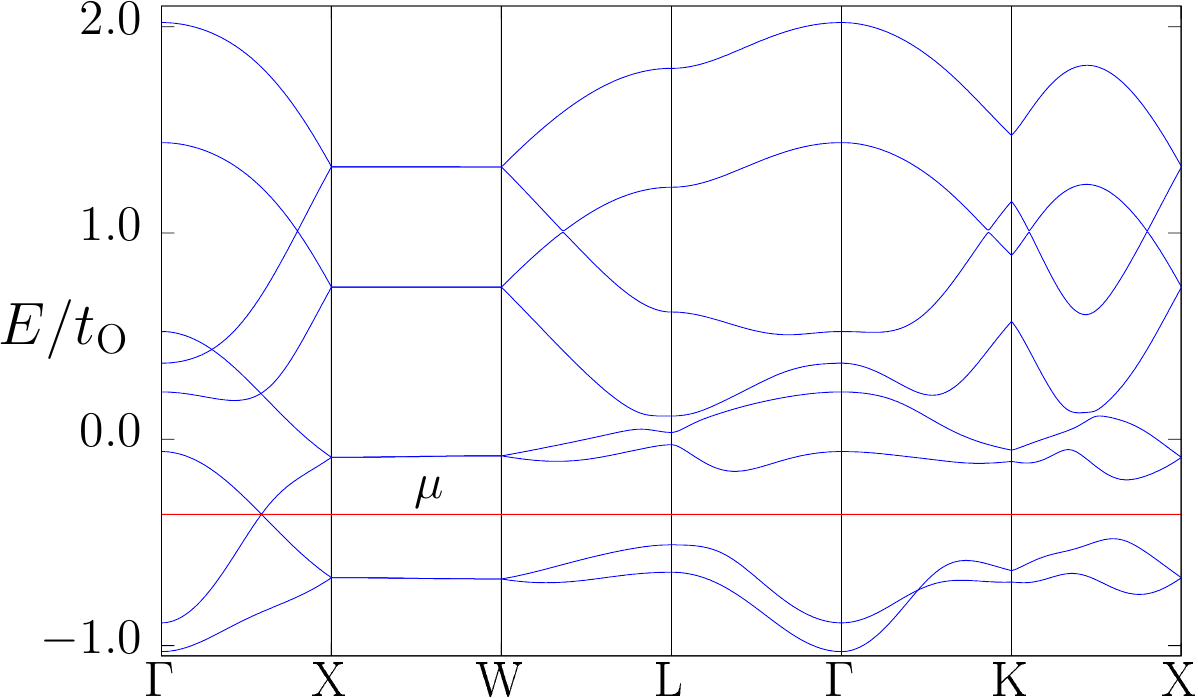}
\end{center}
\caption{Band structure along high-symmetry directions in the fcc Brillouin zone for the r2I2O state at $t_\sigma=-0.9$, $U = 1.3$, and $\rho=0.5$. The r2I2O order breaks the cubic symmetry. The axis $\Gamma$X in the plot is the one distinguished by the order, i.e., the (100) direction for the solution in Eqs.\ (\ref{dope.r2I2O.m1})--(\ref{dope.r2I2O.m4}). The horizontal red line marks the chemical potential.}
\label{fig.r2I2O.bands}
\end{figure}

For $\rho>0.5$, the system returns to the r2I2O state.
Interestingly, the CDW~I and r2I2O states that meet close to $\rho=0.5$ are both Weyl semimetals with two Weyl points. These Weyl points occur at the Fermi energy for quarter filling, i.e., they involve a different pair of bands compared to the TWS close to half filling. As an example, we plot in Fig.\ \ref{fig.r2I2O.bands} the band structure for $t_\sigma=-0.9$, $U = 1.3$, and $\rho=0.5$, for which the r2I2O state is slightly favored over CDW~I. The Weyl points are located on the (100) ($k_x$) axis. These phases are interesting since two is the smallest possible number \cite{NiN81} and since they should show a large anomalous Hall effect \cite{YLR11}. (A multilayer system with two Weyl points has been analyzed by Burkov and Balents \cite{BuB11}).
The mechanism favoring a Weyl state around $\rho=0.5$ is analogous to the one effective close to half filling, which is discussed below.

At even larger $\rho$, the r2I2O order gives way to CDW~I again. At large $U$, this phase surrounds a new phase, denoted by CDW~III. It is the most complicated phase we have encountered, characterized by the spins
\begin{align}
\mathbf{m}_1 &= (a,b,c) , \\
\mathbf{m}_2 &= (a,-c,-b) , \\
\mathbf{m}_3 &= (d,e,-e) , \\
\mathbf{m}_4 &= (f,g,-g)
\end{align}
and only the occupations $n_1$ and $n_2$ being equal. In the specific example, $\mathbf{m}_1$ and $\mathbf{m}_2$ are related by mirror symmetry with respect to the (011) plane and the other two spins are parallel to the same plane.
The degeneracy is $24$-fold, resulting from a factor of $6$ for selecting the pair of equally occupied sites, a factor of $2$ from twofold rotation interchanging the two other sites, and a factor of $2$ from inverting all spins.
Interestingly, all other orders are obtained as special cases: rAIAO for $c=b$, $d=f=-a$, and $e=-g=b$ (AIAO is a special case of rAIAO); CDW~I with sites interchanged compared to Eqs.\ (\ref{dope.CDW.I.m1})--(\ref{dope.CDW.I.m4}) for $b=a$, $d=-c$, $e=a$, and $g=f$; CDW~II for $a=0$, $c=-b$, $f=-d$, and $g=e$; and r2I2O for $c=b$, $d=f=a$, and $e=-g=-b$.

All magnetic phases discussed so far in this section are ferrimagnetic. The broad region of ferrimagnetic states is due to a tendency towards ferromagnetism caused by the Stoner mechanism \cite{Maj00}, modified by magnetic anisotropies due to strong spin-orbit coupling.
We note that Shinaoka \textit{et al.}\ \cite{SMM13} have obtained various such phases for $A_2\mathrm{Mo}_2\mathrm{O}_7$, where $A$ is a rate-earth element, within \textit{ab-initio} calculations and discussed them based on the interplay of anisotropic exchange, Dzyaloshisky-Moriya interaction, and local Ising anisotropy.
The transitions in the series CDW~I--r2I2O--CDW~II--CDW~I--r2I2O--CDW~I--CDW~III--CDW~I are all of first order with discontinuous changes of the magnetization direction.

For increasing $\rho$, another first-order transition leads towards antiferromagnetic AIAO order. Only at small $U$, a region without magnetic or charge order intervenes.
The appearance of antiferromagnetic order close to half filling can be understood from a standard perturbative treatment of electron hopping, leading to antiferromagnetic kinetic exchange \cite{And59}. The band structure changes from gapped at large $U$ to a TWS for smaller $U$. Except close to half filling, the AIAO-TWS transition hardly depends on the filling down to $\rho\approx 0.75$ but is instead only controlled by the interaction $U$. This is surprising since for the heavy doping corresponding to $\rho\approx 0.75$, the Weyl points are far above the Fermi energy and their formation should have negligible effect on the free energy. We conclude that it is actually the AIAO magnetic order that is energetically favored in this range, while the TWS is just a secondary consequence of that order. If the interaction $U$ and thus the spin polarizations $\mathbf{m}_i$ are small, the ordering splits the quadratic band-touching point of the nonmagnetic SM phase into eight Weyl points. For $U\approx 1.3$, these Weyl points are shifted to the L points and annihilate, leading to the insulating AIAO phase.

Finally, the AIAO order gives way to a nonmagnetic phase. A very narrow TWS phase intervenes between the gapped AIAO state and the nonmagnetic state also at larger $U$, which is not resolved in Fig.\ \ref{fig.PDcut1}. The spin polarization decreases to zero continuously, though very rapidly, as a function of $\rho$.

The critical interaction $U$ for the transition between the AIAO and TWS phases and the nonmagnetic phase generally increases with $\rho$, as explained in Appendix \ref{app.phasym}. Close to half filling ($\rho=1$), the stability of the TWS is enhanced, leading to the nonmonotonic dependence of the critical $U$ on $\rho$ seen in Fig.\ \ref{fig.PDcut1}. The reason is that splitting the quadratic $\Gamma_8$ band-touching point into Weyl points reduces the density of states close to the Fermi energy and thus lowers the free energy. As mentioned in the introduction, this mechanism is most efficient if the Fermi energy coincides with the energy of the $\Gamma_8$ point, i.e., at half filling.

\begin{figure}[htb]
\begin{center}
\includegraphics[width=\columnwidth]{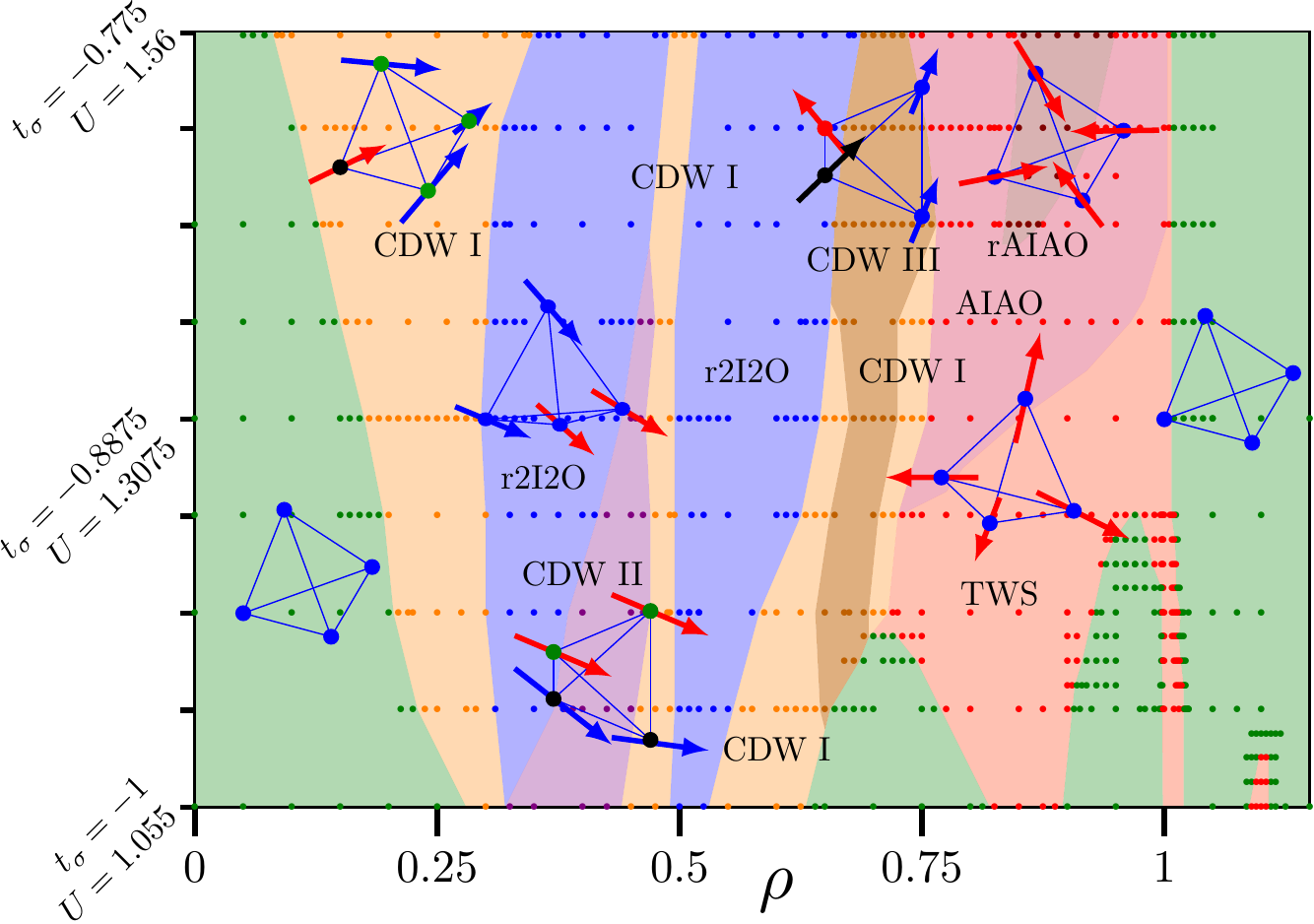}
\end{center}
\caption{Mean-field phase diagram for varying $t_\sigma$ and $U$ along the orange line with dots in Fig.\ \ref{fig.PD1} and also varying $\rho$. The considered parameter points are shown as colored dots. The regions shaded in less saturated colors represent our best guess at the regions occupied by the various phases. The magnetic and charge order in the various phases is indicated in the insets. The acronyms are explained in the text.}
\label{fig.PDcut2}
\end{figure}

We briefly consider another cut through the $(t_\sigma,U,\rho)$ phase diagram, which is indicated by the orange line with dots in Fig.\ \ref{fig.PD1}. The phase diagram is shown in Fig.\ \ref{fig.PDcut2}. Overall, the phase diagram is similar to the one in Fig.\ \ref{fig.PDcut1}. The extent of some phases, notably CDW~II and CDW~III, has expanded but the general sequence of phases is the same. There is now an island of a gapped system with rotated all-in-all-out (rAIAO) order surrounded by the gapped AIAO phase. The rotation angle $\Phi$ is smaller than for the rAIAO phase at $\rho=1$. The transition between AIAO and rAIAO is of first order.
The nonmonotonic dependence of the transition between the TWS and nonmagnetic states on $\rho$ close to half filling is much more pronounced than in Fig.~\ref{fig.PDcut1}---the stabilization of the TWS phase at $\rho\approx 1$ is strongly enhanced. This is mainly due to the cut being nearly parallel to the TWS-to-nonmagnetic transition in Fig.\ \ref{fig.PD1}.
Moreover, the transition between the AIAO and TWS phases is not flat as in Fig.\ \ref{fig.PDcut1}, but note that now both $U$ and $t_\sigma$ vary along the vertical axis. The narrow TWS phase between the AIAO and nonmagnetic states is resolved in Fig.~\ref{fig.PDcut2}.

\subsection{Search for broken translational symmetry}
\label{sec.DW}

In this section, the stability of the Weyl state of the undoped and weakly doped pyrochlore system with respect to breaking the translational symmetry is assessed. We have searched for charge and spin DWs and combinations of the two. As discussed above, breaking of translational symmetry leads to backfolding of bands. Weyl points of opposite chirality that are far apart in the original Brillouin zone can end up close together. A small change of the order parameters could then move them to the same $\mathbf{k}$ point, where they can gap out, which could reduce the free energy \cite{WCA12,WaZ13,WaY16,CTS16,LPT16}. Since bands close to Weyl points of opposite chirality would hybridize in such a state, it would form an axion insulator \cite{WaZ13}.
Note that Sec.\ \ref{sec.doping} demonstrates that inhomogeneous charge configurations can occur.

In the absence of a DW, the Weyl points form the corners of a cube with neighboring points having opposite chirality, see Fig.\ \ref{fig.WP1}. The optimal situation for the described mechanism is a DW in the (001) or equivalent directions. The backfolding can then place all Weyl points close together in pairs of opposite chirality. For a DW in the (111) or equivalent directions, only two Weyl points of opposite chirality can end up close together, namely the two with their separation vector parallel to the ordering vector $\mathbf{Q}$ of the DW.
To allow for DW solutions, we consider supercells consisting of $n$ primitive unit cells such that the translational symmetry is reduced in the (001) or (111) direction. The average occupations $n_i$ and spin polarizations $\mathbf{m}_i$ are allowed to be distinct for all $4n$ sites in the supercell.

We have performed iterative mean-field calculations for DWs along (001) with $n=3,4,5$ and along (111) with $n=2,3,4,5$, for various values of $t_\sigma$ and $U$. We have started from up to 30 sets of random initial values for all mean-field parameters $n_i$ and $\mathbf{m}_i$ for each case. Furthermore, we have also initialized the iteration with DW states imposed by hand that are discussed below. No DW states breaking translational symmetry have been found. Under iteration, all spin polarizations flow to the AIAO order and all charges to uniform occupation. Even if the locations of the Weyl points in the uniform state are fine tuned, by changing $U$ and $t_\sigma$, to fold back on top of each other, we do not obtain a stable DW. This has been tested for DWs along (001) with $n=3,4$ and along (111) with $n=2,3,4$. To check whether small shifts of the Fermi energy have any effect, we have varied $\rho$ between $0.98$ and $1.02$ in all cases. In the remainder of this section, we discuss why the TWS state is robust.

\begin{figure}[htb]
\begin{center}
\raisebox{1ex}{(a)}\includegraphics[scale=0.65]{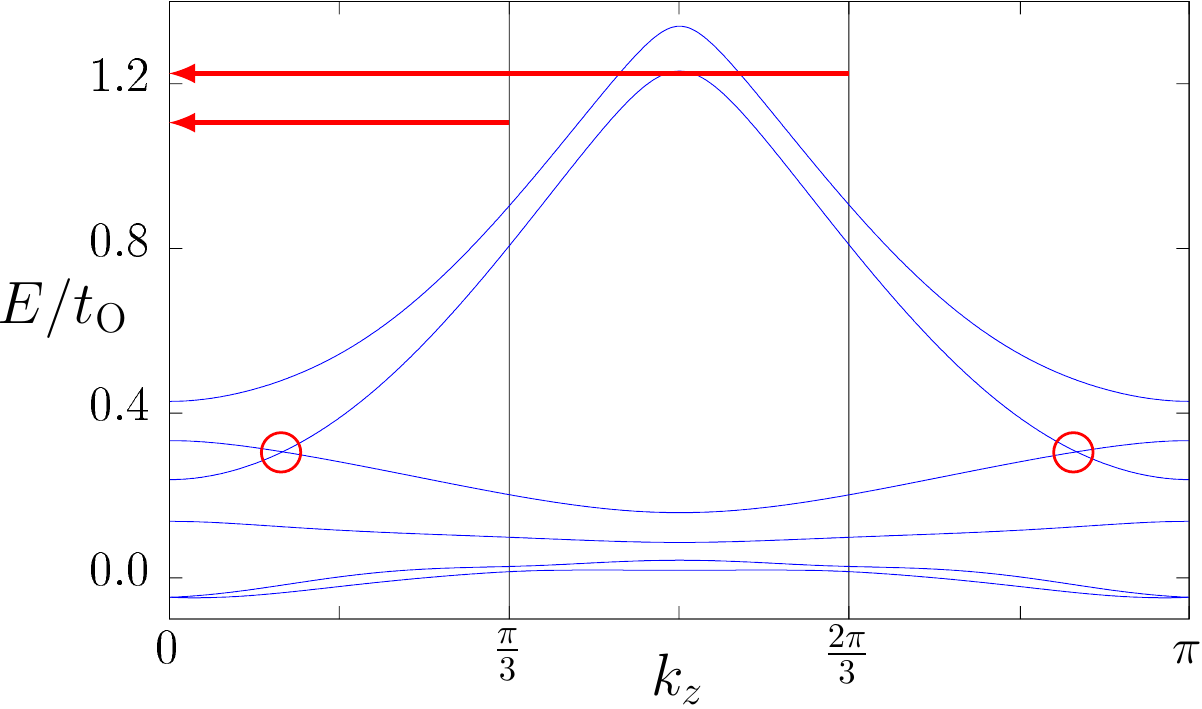}\\[2ex]%
\raisebox{1ex}{(b)}\includegraphics[scale=0.65]{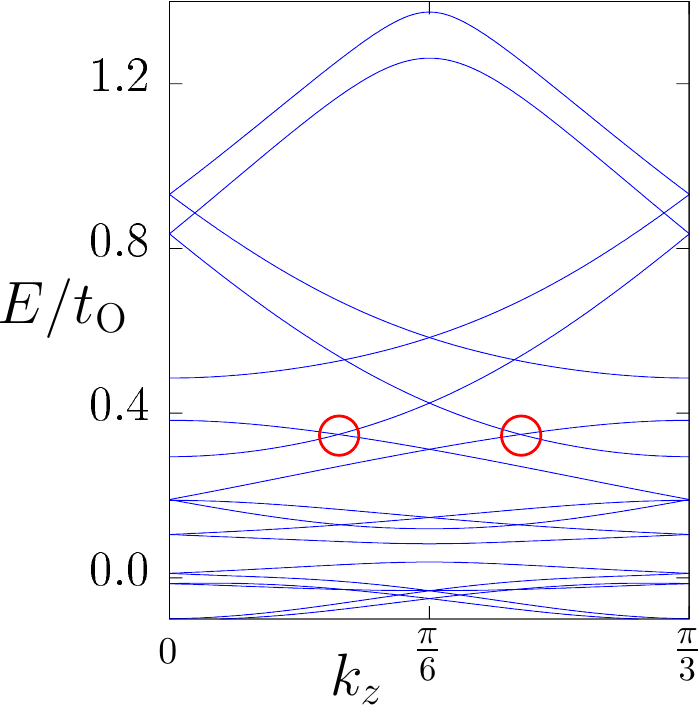}
\end{center}
\caption{(a) Band structure along the (001) direction for the TWS with $t_\sigma=-0.775$ and $U=1.515$. The two Weyl points are marked by red circles. When a supercell consisting of $n=3$ primitive unit cells is considered, the Brillouin zone is backfolded and the bands are shifted into the smaller Brillouin zone as indicated by the red arrows. (b) The resulting band structure for the supercell with $n=3$.}
\label{fig.DW001.bands}
\end{figure}

We first demonstrate that the Weyl points can indeed be moved and induced to annihilate by imposing a DW. We take parameters $t_\sigma=-0.775$ and $U=1.515$, which lie within the TWS phase in Fig.\ \ref{fig.PD1}. The band structure along the $k_z$ direction through a pair of Weyl points is shown in Fig.\ \ref{fig.DW001.bands}(a). It is of course possible to describe this situation using a larger unit cell. For the example of $n=3$ primitive unit cells stacked in the (001) direction, the bands are shifted as indicated by the arrows in Fig.\ \ref{fig.DW001.bands}(a), which leads to the band structure in Fig.\ \ref{fig.DW001.bands}(b). Now, a CDW is switched on by adding
\begin{equation}
H_\mathrm{CDW} = -U\delta \sum_{\mathbf{r}} \cos\left(\frac{\pi}{n}\, \mathbf{r}
  \cdot \hat{\mathbf{z}}\right) \sum_\sigma c^\dagger_{\mathbf{r}\sigma} c_{\mathbf{r}\sigma}
\label{CDW.type.5}
\end{equation}
to the Hamiltonian. Here, $\mathbf{r}$ traverses all sites of the pyrochlore structure and $\hat{\mathbf{z}}$ is the unit vector in the (001) direction. In a selfconsistent calculation, $H_\mathrm{CDW}$ would represent the Hartree potential resulting from the CDW. Translational symmetry is broken for $n=2,3,\ldots$

\begin{figure}[htb]
\begin{center}
\raisebox{1ex}{(a)}\includegraphics[width=0.8\columnwidth]{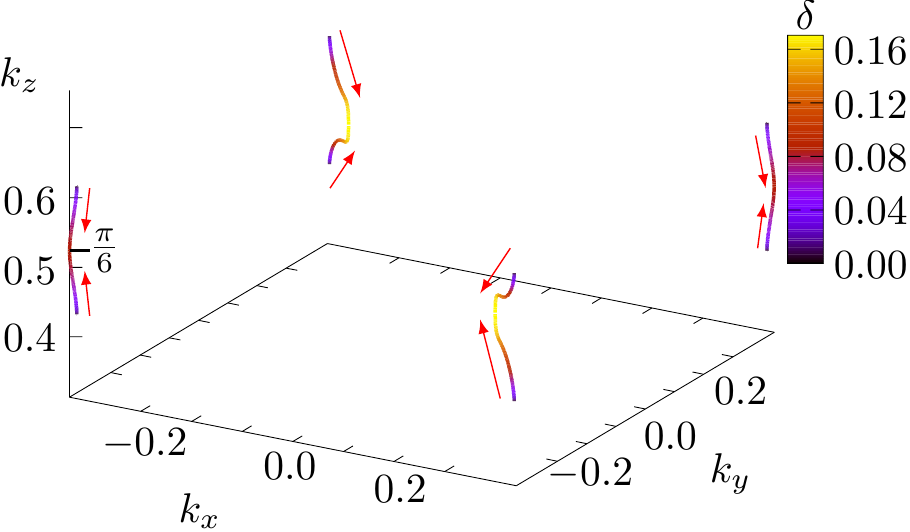}\\[1ex]%
\raisebox{1ex}{(b)}\includegraphics[width=0.8\columnwidth]{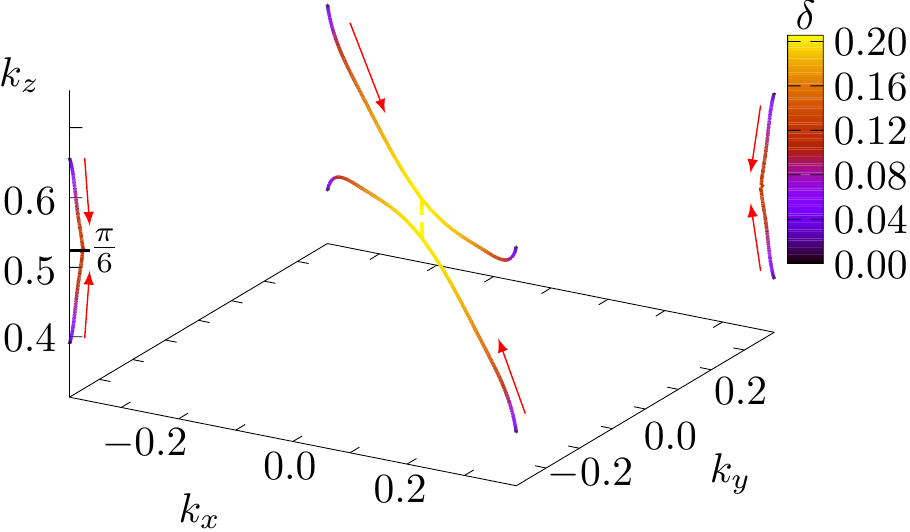}\\[1ex]%
\raisebox{1ex}{(c)}\includegraphics[width=0.8\columnwidth]{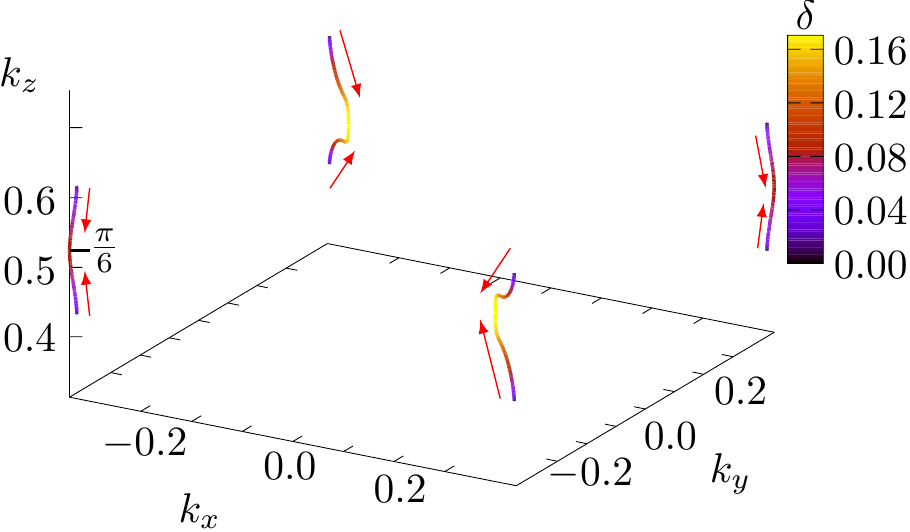}
\end{center}
\caption{Shift of Weyl points in the backfolded Brillouin zone for a CDW in the (001) direction with $n=3$, $t_\sigma=-0.775$, and (a) $U=1.535$, (b) $U=1.525$, and (c) $U=1.515$. The colored lines denote the paths taken by the Weyl points for increasing CDW amplitude $\delta$.}
\label{fig.CDW001.move}
\end{figure}

The band structures in Fig.\ \ref{fig.DW001.bands} correspond to a vanishing CDW amplitude, $\delta=0$. For $\delta\neq 0$, the band energies change and energy gaps open at the new zone boundaries. The band touching at the Weyl points cannot  be gapped out since these points are topologically protected by a nonzero Chern number. Hence, it can only change if Weyl points are annihilated or created in pairs.
Figure \ref{fig.CDW001.move} illustrates how the Weyl points move for increasing amplitude $\delta$. We find that for sufficiently large $\delta$, the Weyl points always meet and gap out. However, it is not always the case that the Weyl points that start out close together annihilate. This simple scenario is only realized in Fig.\ \ref{fig.CDW001.move}(a). For smaller $U$, Figs.\ \ref{fig.CDW001.move}(b) and \ref{fig.CDW001.move}(c), the Weyl points are further apart for $\delta=0$ and the annihilation is more complex. 

A closer look at Fig.\ \ref{fig.CDW001.move} reveals that the Weyl points do not all annihilate at the same value of $\delta$. Instead, two pairs annihilate for much smaller $\delta$ than the other two. This is consistent with symmetry since the CDW reduces the magnetic point group to $D_{2h}(C_{2h})$, which contains a twofold rotation about the \textit{z}-axis but no fourfold rotations.
$D_{2h}(C_{2h})$ also does not contain threefold rotation axes so that the Weyl points are not pinned to the (111) direction.

Since a supercell consisting of $n$ primitive unit cells stacked in the (001) direction comprises $2n$ rectangular layers, CDWs with half integer $n=3/2,5/2,\ldots$ in Eq.\ (\ref{CDW.type.5}) are possible. Consecutive chains of maximally occupied sites are then rotated by $90^\circ$ with respect to each other and the supercell contains $2n$ primitive unit cells. In this case, the magnetic point group is $D_{4h}(D_{2h})$, which retains fourfold rotation combined with time reversal. Consequently, all eight Weyl points are linked by symmetries. However, the threefold rotation symmetries are still broken.
This case is included in our unrestricted Hartree-Fock calculations. If we choose a supercell containing a number $m$ of primitive unit cells and $m$ is odd, the mean-field equations are free to select a DW with integer $n=m$ or with half-integer $n=m/2$.

\begin{figure}[htb]
\begin{center}
\includegraphics[width=0.8\columnwidth]{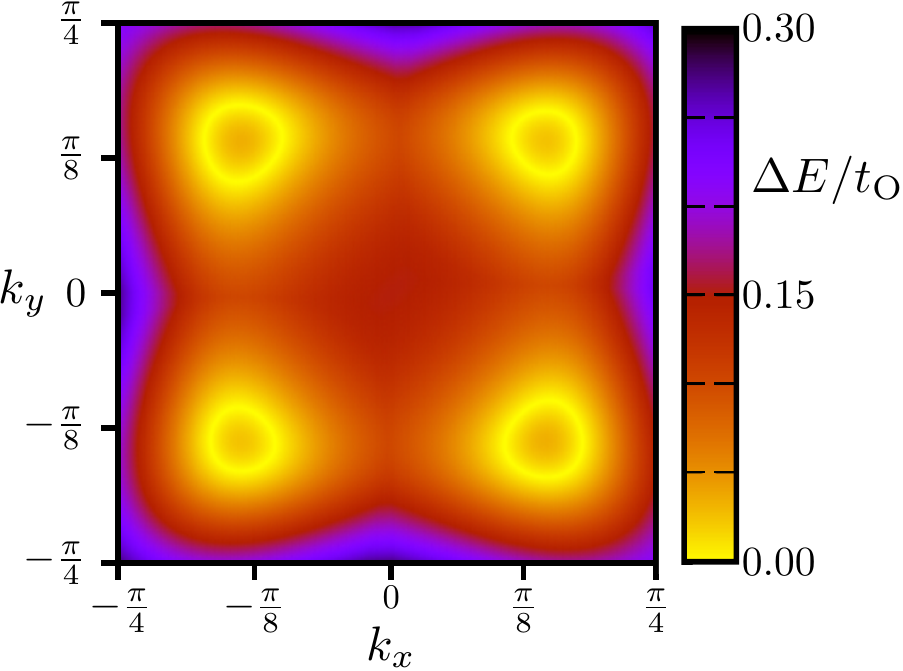}
\end{center}
\caption{Energy difference $\Delta E$ between the two bands that for smaller $\delta$ form the Weyl points for a CDW state with $t_\sigma=-0.775$, $U=1.515$, $n=4$, and $\delta=0.8$, in the high-symmetry plane $k_z=\pi/8$.}
\label{fig.CDW001.loops}
\end{figure}

If the Weyl points are lying close together in the backfolded Brillouin zone, Fig.\ \ref{fig.CDW001.move}(a), we find that typically a small amplitude $\delta$ is sufficient to annihilate them. However, this does not lead to the opening of a global gap. Rather, we find a semimetal with Weyl rings, i.e., one-dimensional loops of crossings of otherwise nondegenerate bands.
Although this state ultimately does not help to understand the stability of the TWS, it is worth to briefly discuss it as an alternative topological state. The Weyl loops are lying in the $k_z=\pi/2n$ plane, which forms a face of the backfolded Brillouin zone. They are shown in Fig.\ \ref{fig.CDW001.loops}. The $k_z=\pi/2n$ plane is invariant under the mirror reflection $m_z$, which is the only nontrivial symmetry in the little magnetic group $C_{1h}$ of that plane. The two touching bands transform according to different irreps of $C_{1h}$ so that crossings are allowed by symmetry. The crossing of two bands in a two-dimensional manifold is generically one dimensional. Away from the high-symmetry plane, the little group is trivial ($C_1$) and all band crossings are avoided. 
The Weyl rings can also be understood based on the theory of symmetry-protected topological states \cite{ChS14,CTS16}.
Weyl rings protected by a mirror symmetry have been found by band-structure calculations for $\mathrm{GdPtBi}$ in a magnetic field \cite{HKW16}.
The origin of the loops is distinct from the case discussed in Ref.\ \cite{WGM17} for pyrochlore iridates, where they are due to an approximate symmetry.
The appearance of Weyl rings is not favorable for the formation of DWs since they increase the density of states close to the Fermi energy, even if the band structure is fine tuned so that the loop appears at the Fermi energy. Hence, DWs that break the mirror symmetry generally have a lower free energy.

For CDWs in the (111) direction, a supercell is formed by stacking $n$ primitive unit cells along (111). A CDW with $n=1$ corresponds to a case where the alternating triangular and kagome layers have different occupation numbers. This is the CDW~I phase introduced in Sec.\ \ref{sec.doping}. For $n\ge 2$, the Brillouin zone is backfolded. As noted above, only the two Weyl points with connecting vector parallel to (111) can end up close together. We find that they annihilate and gap out for sufficiently large $\delta$. The CDW breaks lattice symmetries in such a way that these two Weyl points are no longer related to the other six by symmetry.
CDWs along (111) do not remove the threefold rotation symmetry about this direction. Thus the Weyl points that can annihilate are pinned to the (111) direction.

The most favorable situation for DW formation is realized when Weyl points of opposite chirality are mapped to the same point in the backfolded Brillouin zone. Then an infinitesimal DW amplitude should be able to gap them out. As noted above, we have not obtained stable DW solutions even in that case. To understand why not, we consider the deviations $\Delta n_i$ and $\Delta\mathbf{m}_i$ of $n_i$ and $\mathbf{m}_i$, respectively, from the TWS state as small perturbations.
The free energy does not contain a term of first order in $\Delta n_i$ and $\Delta\mathbf{m}_i$. For the occupations, the first-order contribution to the mean-field energy is proportional to $\sum_i n_i^{(0)}\, \Delta n_i$, where $n_i^{(0)}$ is the unperturbed occupation number at site $i$. However, these are uniform in the TWS state so that the contribution is proportional to $\sum_i \Delta n_i$. This sum vanishes since the number of electrons is fixed.
For the spin polarizations, the first-order correction to the energy is proportional to $\sum_i \mathbf{m}_i^{(0)} \cdot \Delta\mathbf{m}_i$, where $\mathbf{m}_i^{(0)}$ is the unperturbed spin polarization \cite{endnote.Jtensor}. Hence, only the component $\Delta m_i \equiv (\mathbf{m}_i^{(0)}/|\mathbf{m}_i^{(0)}|) \cdot \Delta\mathbf{m}_i$ parallel to $\mathbf{m}_i^{(0)}$ contributes. Since the unperturbed $\mathbf{m}_i^{(0)}$ have uniform magnitude, the first-order correction to the energy is proportional to $\sum_i \Delta m_i$. The $\Delta m_i$ can now be decomposed into a uniform contribution $\overline{\Delta m_i}$, where the average is over all sites $i$, and a site-dependent one, $\Delta m_i-\overline{\Delta m_i}$. We already know that the TWS state is stationary for uniform perturbations so that the free energy cannot depend linearly on $\overline{\Delta m_i}$. We can thus subtract the uniform part from $\Delta m_i$ without changing the free energy to linear order. But then we obtain $\sum_i \Delta m_i=0$ and the magnetic contribution is found to vanish to linear order.

Due to the vanishing of the first-order contributions, the uniform TWS state remains a stationary point of the free energy when DWs are allowed. We first consider the stability of this stationary point understood as a fixed point of the iterative mapping described by the mean-field equations. The space of perturbations $\Delta n_i$ and $\Delta\mathbf{m}_i$ has $16n-1$ independent dimensions, see Sec.\ \ref{sec.model}. If there were an unstable direction, a random set of $\Delta n_i$ and $\Delta\mathbf{m}_i$ would have nonzero overlap with it with probability one. In this sense, it is sufficient to check the stability for a single random set of starting values. However, if the initial overlap is very small, the iteration might not pick up the unstable direction during a feasible number of iterations. For this reason, we have always taken several sets of random starting values. Since they all converge to the uniform stationary point, we conclude that all directions are stable. It remains to understand the reason for this.

A DW should be most strongly stabilized when it opens a global gap opens at the Fermi energy. This can indeed happen: for $t_\sigma=-0.775$, $U=1.55$, and $n=3$, we have considered $10^5$ sets of small random deviations and find a global gap in $8.8\%$ of cases. Complete splitting between the bands without a global gap is found with a frequency of $13.9\%$ and persisting Weyl points with $77.3\%$.
We will now give an argument as to why a DW is not stabilized even for a global gap.

The stability analysis of the uniform state is not straightforward since the stationary point is a minimum as a function of the spin polarizations $\mathbf{m}_i$ but is a maximum as a function of the occupations $n_i$ \cite{LaW97}. This can be understood based on minimization of the grand-canonical potential or within the functional-integral approach \cite{LaW97,endnote.LaW}. The correct extremization procedure is to first maximize the free energy for fixed $\Delta\mathbf{m}_i$ as a function of the $\Delta n_i$ satisfying $\sum_i \Delta n_i=0$. Let us denote the result by $\mathcal{F}(\Delta\mathbf{m}_1,\Delta\mathbf{m}_2,\ldots)$. Then, $\mathcal{F}$ is minimized with respect to the $\Delta\mathbf{m}_i$. This means that standard stability analysis applies to $\mathcal{F}(\Delta\mathbf{m}_1,\Delta\mathbf{m}_2,\ldots)$.

Let us denote the gap by $2\Delta$, i.e., occupied states close to the Weyl points are shifted downward in energy by $\Delta$. The momentum-space volume that is affected should scale with $\Delta^3$ since the unperturbed dispersion is approximately linear. The typical energy gain of occupied states within this volume is proportional to $-\Delta$. The total gain in the free energy $\mathcal{F}$ should thus scale as $-\Delta^4$. As noted above, $\Delta$ is of first order in the perturbation, i.e., in an overall scaling factor $\lambda$ for all $\Delta\mathbf{m}_i$. This gives a negative contribution of order $-\lambda^4$ to the free energy. However, there is another contribution from the decoupling term in the mean-field ansatz, $NU\sum_i \mathbf{m}_i\cdot\mathbf{m}_i$. This term is positive and scales as $\lambda^2$. Hence, the uniform TWS stationary point is stable against small perturbation.
It is instructive to contrast the case that a normal Fermi surface with Fermi wave number $k_F$ is gapped out. In that case, the affected $\mathbf{k}$-space volume scales as $k_F^2\,\Delta$ and the total energy gain as $-k_F^2\,\Delta^2$ and thus as $-\lambda^2$. This energy can compete with the quadratic decoupling term. This is the situation for the instability of a compensated metal towards an excitonic insulator \cite{exc1,exc4,ZTB11,endnote.BCS}.

A few comments are in order. Our argument does not exclude a first-order transition to a state with large $\Delta\mathbf{m}_i$. We have not found any such state, though. Furthermore, depending on the details of the model, there likely exists a second-order contribution to the total energy gain due to small shifts of the (large) energies of states far from the Weyl points. However, the coefficient of the corresponding quadratic term is expected to be small since it describes the effect of states deep in the Fermi sea. If so, it cannot overcome the large, positive decoupling term.
In other words, the TWS is rather stable against DW formation since the occupied density of states close to the Fermi energy is small to start with. Thus changes in the electronic dispersion at the mean-field level cannot lead to a large reduction that could overcome the positive decoupling term.

\section{Summary and outlook}
\label{sec.summary}

A Hubbard-type model on the pyrochlore lattice has been studied, motivated by the proposed Weyl-semimetal phase in pyrochlore iridates \cite{WTV11,WiK12,ChH12}. For the stoichiometric material, corresponding to half filling, we essentially reproduce the phase diagram of Ref.\ \cite{WGK13}. For large direct iridium-iridium hopping $|t_\sigma|$ and small interaction $U$, we find additional phases, which are identified as type-II Weyl semimetals \cite{SGW15}.
We recall that our parameter $U$ is the ratio of the Hubbard repulsion and the hopping amplitude via oxygen, which can be tuned by hydrostatic pressure. In view of existing experiments on $\mathrm{Eu_2Ir_2O_7}$ under pressure \cite{TIM12}, it is promising to use pressure to tune one of the iridates into the Weyl phase.

Upon hole doping, a plethora of ordered phases appears, which are distinguished by their magnetic order. All of them are noncollinear ferrimagnets. The large number of phases can be attributed to the strong magnetic frustration resulting from the spatially varying easy axes. Some of the phases also show charge order, i.e., different occupations of the four sites in the basic tetrahedron. The various spin and charge orders break lattice symmetries but do not enlarge the unit cell.

The contribution of Weyl points to the free energy is found to be small except close to half filling. Instead, the magnetic state is selected by free-energy minimization and the TWS is a secondary effect occurring for not too strong all-in-all-out magnetic order. This changes close to half filling, where the TWS phase is stabilized since it has lower density of states close to the Fermi energy than the semimetallic or metallic phases.
The all-in-all-out magnetic order, be it in a TWS or gapped state, is destroyed by doping. However, its fragility is highly asymmetric for hole vs.\ electron doping; electron doping destroys the order much more rapidly because of the difference between the effective masses in the valence and conduction bands.

Close to quarter filling, the ordered phases have a TWS band structure with two Weyl points. These Weyl points appear at the Fermi energy (for quarter filling) and thus involve different bands than the Weyl points close to half filling. A realistic description of pyrochlore systems in this heavily doped regime would likely require additional bands to be taken into account. Coexisting conventional Fermi surfaces would not rule out the existence of Weyl points, in principle, but would shift the Fermi energy and thus the doping levels at which they occur.

It would be straightforward to obtain the temperature dependence of the new phases within Hartree-Fock theory. At half filling, this has been done in Ref.\ \cite{WGK13}, where the TWS phase is found to be thermally much less stable than the insulating AIAO and rAIAO phases. As noted there, it is very reasonable that the gapped insulating phases are thermally more stable than the gapless Weyl state. We thus expect the additional Weyl states close to quarter filling to have low critical temperatures.

We have also searched for DWs that break translational symmetry at and close to half filling. This was motivated by the expectation that Weyl points of opposite chirality that are placed close together by backfolding could merge and gap out. We have found the TWS state with unbroken translational symmetry to be stable even for the most critical situation where the Weyl points end up on top of each other. This can be understood based on the low density of states close to the Fermi energy in the TWS phase, which gives little room for a further reduction of the free energy. A simple harmonic CDW is additionally disfavored by symmetry-allowed crossings of bands that prevent the opening of a gap.

The main limitation of our approach is the omission of fluctuations. Since we consider the low-temperature limit, thermal fluctuations are not important. However, the Hubbard interaction for which the interesting phases occur is on the order of the band width. The model is thus in the intermediate-coupling regime. This could lead to Mott physics \cite{GWJ12}, which is not captured by the Hartree-Fock approach. In this approximation, the gapped AIAO phase turns into a metal away from half filling since the Fermi energy lies in a band, whereas a Mott insulator would survive in a finite doping range.
Furthermore, in the limit of pure spin models, the antiferromagnetic and ferrimagnetic states show quantum fluctuations. Such fluctuations are present also in the metallic and semimetallic phases but should be relatively benign since the system is three dimensional. Their leading effect if correlations are not too strong should be to reduce the spin polarizations relative to the mean-field prediction.
It would be of interest to study these effects beyond the Hartree-Fock level, but mainly if a candidate pyrochlore system with TWS or one of the other magnetic metallic and semimetallic phases would be found.
Our general results---the likely existence of a sequence of complicated magnetic orders in doped pyrochlores and the robustness of the TWS state at half filling against DW formation---are expected to survive for moderately strong correlations.

Concerning the question of the lack of experimental realizations of pyrochlore TWSs, we conclude that an instability towards a DW state cannot explain this observation. Also, weak unintentional doping cannot explain it since weak hole doping preserves the TWS state while electron doping quickly leads to a nonmagnetic state, which is not observed. Instead, the absence of TWS state is likely due to the on-site repulsion exceeding the hopping-dependent critical value so that the materials end up in the gapped AIAO phase.

\acknowledgments

The authors wish to thank M. Breitkreiz, C. Ko\-schenz, and W. Witczak-Krempa for fruitful discussions and T. Ludwig for help with the computations. Financial support by the Deut\-sche For\-schungs\-ge\-mein\-schaft through project A04 of Collaborative Research Center SFB 1143, is gratefully acknowledged.

\appendix

\section{Particle-hole asymmetry}
\label{app.phasym}

The phase diagrams in Figs.\ \ref{fig.PDcut1} and \ref{fig.PDcut2} are highly particle-hole asymmetric relative to half filling ($\rho=1$). Here, we give a more detailed explanation for this asymmetry. We start from the SM phase with quadratic band-touching point at $\Gamma$. Let $2\Delta$ be the full gap in the AIAO phase, i.e., valence (conduction) band states at $\Gamma$ are shifted down (up) in energy by $\Delta$. We approximate the dispersion by
\begin{equation}
E = \pm \sqrt{(k^2/2m_\pm)^2 + \Delta^2} ,
\end{equation}
where the sign distinguishes the two bands and $m_+$ ($m_-$) is the effective mass in the conduction (valence) band. We assume electron doping and first consider the case that the chemical potential is significantly larger than $\Delta$, i.e., far above the gap. Then the radii of the regions in momentum space which are affected by the gap are $k_\pm \approx \sqrt{2m_\pm \Delta}$ for the two bands. The typical energy shift for the affected states scales with $\Delta$ and is positive (negative) for the conduction (valence) band. The energy difference between the gapped and ungapped phases then scales as
\begin{equation}
\Delta E \sim (m_+^{3/2}-m_-^{3/2})\, \Delta^{5/2} .
\end{equation}
If the effective mass in the conduction band is larger, $\Delta E$ is positive and the gapped AIAO phase is unfavorable.

In the opposite limit, $\mu\lesssim \Delta$, the radii of regions affected by the gap are unchanged but the chemical potential now lies close to the (very flat) bottom of the conduction band. Not all states in the conduction band that are affected by the gap are occupied but only those with $k \lesssim \sqrt{2m_+\mu}$. Only the occupied states contribute to the energy so that the energy difference now scales as
\begin{equation}
\Delta E \sim (m_+^{3/2}\mu^{3/2}-m_-^{3/2}\Delta^{3/2})\,\Delta .
\end{equation}
This difference becomes negative for $\Delta > (m_+/m_-)\,\mu$. Hence, the AIAO phase can be stabilized by making $\Delta$ sufficiently large, which requires a large $U$. For this reason, the critical $U$ grows rapidly for increasing $\rho$.


\begin{thebibliography}{99}
\bibitem{endnote.Jtensor}The magnetic energy involves a site-dependent exchange tensor $J_i$. However, for all-in-all-out order, the spin polarizations $\mathbf{m}_i^{(0)}$ point along the (111) and equivalent directions and are of equal magnitude. Since the system has threefold rotation symmetry about this axis, it must be a principal axis of $J_i$ and the other two principal axes are orthogonal to it but otherwise arbitrary. Therefore, the mean-field energy contribution is proportional to $\mathbf{m}_i^{(0)}\cdot \Delta\mathbf{m}_i$.
\bibitem{endnote.LaW}In the functional-integral approach, this follows from the charge part of the Hubbard interaction being repulsive, whereas the spin part is attractive. The fluctuations in the Hubbard-Stratonovich field for the former are orthogonal in the complex plane to variations of the real charge order parameter. The action has a minimum as a function of the fluctuations but a maximum as a function of real charge variations.
\end{thebibliography}
\end{document}